%% file: main.tex
\documentclass[conference]{IEEEtran}
\usepackage[T1]{fontenc}
\IEEEoverridecommandlockouts
\usepackage{cite}
\usepackage{amsmath,amssymb,amsfonts}
\usepackage{graphicx}
\usepackage{textcomp}
\usepackage{xcolor}
\usepackage{booktabs}
\usepackage{url}
\usepackage{xspace}
\usepackage{eso-pic}
\usepackage{tikz}
\usetikzlibrary{arrows.meta,positioning}
\definecolor{bluF}{HTML}{DAE8FC}\definecolor{bluB}{HTML}{6C8EBF}
\definecolor{grnF}{HTML}{D5E8D4}\definecolor{grnB}{HTML}{82B366}
\definecolor{orgF}{HTML}{FFE6CC}\definecolor{orgB}{HTML}{D79B00}
\definecolor{purF}{HTML}{E1D5E7}\definecolor{purB}{HTML}{9673A6}
\definecolor{yelF}{HTML}{FFF2CC}\definecolor{yelB}{HTML}{D6B656}
\newcommand{\hd}[1]{{\footnotesize\bfseries #1}\\[2pt]}
\newcommand{\mn}[1]{{\ttfamily #1}}
\newcommand{\repoURL}{\url{https://github.com/research-repo-bai/MongoBench}}
\def\BibTeX{{\rm B\kern-.05em{\sc i\kern-.025em b}\kern-.08em
    T\kern-.1667em\lower.7ex\hbox{E}\kern-.125emX}}

\input{gen3/numbers.tex}
\newcommand{\mongod}{\texttt{mongod}\xspace}
\newcommand{\find}{\texttt{find()}\xspace}

\begin{document}

\title{MongoBench: Do LLMs Know What a MongoDB\\Query Will Do?}

\author{\IEEEauthorblockN{Qibai Chen}
\IEEEauthorblockA{\textit{Independent Researcher}\\
United States \\
qibaic@alumni.cmu.edu}
\and
\IEEEauthorblockN{Haoting Qiu}
\IEEEauthorblockA{\textit{Independent Researcher}\\
United States \\
qiu.haot@northeastern.edu}
\and
\IEEEauthorblockN{Yezhou Cheng}
\IEEEauthorblockA{\textit{University of Wisconsin--Madison}\\
United States \\
cheng222@wisc.edu}
\and
\IEEEauthorblockN{Mengxiao Qian}
\IEEEauthorblockA{\textit{Independent Researcher}\\
United States \\
qmxresearch@gmail.com}}

\maketitle

\AddToShipoutPictureFG*{%
  \AtPageLowerLeft{%
    \raisebox{0.40in}[0pt][0pt]{%
      \hspace{0.75in}\normalfont\fontsize{8}{9}\selectfont Accepted at CAIT 2026. Author's accepted manuscript.%
    }%
  }%
}

\begin{abstract}
LLM agents increasingly operate databases on their own: they generate queries and run them against live data.
Two checks are commonly used to judge such an agent---whether it produces a valid query, and whether it reaches the right final answer---but neither tells us whether the agent knows \emph{what its query will actually do} before it runs.
We argue that this \emph{consequence prediction} ability must be evaluated in its own right, and we build \textbf{MongoBench} to measure it: given a collection of documents and a MongoDB \find filter, a model must return the exact set of matching \texttt{\_id}s, with ground truth from a real \mongod server.
The \nTotal{} items each compose three or four documented matching pitfalls and are proposed symmetrically by three vendors' models, so results stratify by item source.
Evaluating four models---frontier systems from three vendors and one open-weight reasoner---accuracy ranges from \accGem\% to \accGpt\%, a \gapTop-point spread.
Each proposer mispredicts 10--31\% of the items it wrote itself, so producing a query does not imply predicting its effect.
A failure-mode analysis shows errors are better predicted by \emph{which} semantic pitfall a query exercises---range comparisons and \texttt{null}---than by surface measures such as document count or the number of composed rules.
Prepending a verified rule sheet helps unevenly; executing the original filter against the engine yields reliable match sets, but delegation must be checked: on \gemAlteredPctMean\% of items (averaged over three runs) one model's executed filters never passed a canonical check against the intended query, and it answered about \gemAlteredWrongMean{} of them wrong, while the other three passed on every item.
We release the benchmark, the \mongod oracle, and the evaluation code with all prompts at \repoURL.
\end{abstract}


\section{Introduction}
\label{sec:intro}

LLM coding agents now resolve real software-engineering tasks with little human oversight~\cite{swebench-2024}, and in doing so increasingly read, write, and execute database queries on their own.
To judge whether such an agent is trustworthy, we usually look at two things: can it produce a syntactically valid query, and does it reach a correct final answer (for instance, after running the query and reading the result).
Both miss a capability that sits between them and is safety-critical in its own right: before an agent runs a query, does it know \emph{which objects the query will affect}?
An agent that emits a valid \texttt{delete} filter need not know which documents that filter removes; an agent that can call a database tool need not pass the tool the query it meant.
Predicting the consequence of an action, and preserving that action's meaning when handing it to an executor, is what lets an agent estimate a blast radius, explain an empty result, or veto a wrong query before damage is done.
This matters most when running the query first is not an option: a destructive or irreversible write, a code review with only the schema or a sample in context, or any step that must judge a query before it touches live data.

We make this \emph{consequence prediction} ability the object of study.
For document stores it has a clean, machine-checkable form: given documents $D$ and a \find filter $Q$, which \texttt{\_id}s does $Q$ match under the engine's rules?
Research on LLMs and document-store queries has instead focused on \emph{generation}---translating natural language to a query string~\cite{sm3-2024,text2nosql-2024}---which does not test whether the model understands the matching semantics of what it writes.
Those semantics are full of documented pitfalls: a scalar condition also matches arrays containing the value; multiple conditions on an array may be met by different elements unless \texttt{\$elemMatch} is used; \texttt{null} also matches missing fields; a range comparison against \texttt{null} matches nothing; and equality with an embedded document is field-order sensitive.
A model that writes a valid filter while mispredicting which documents it affects can, if that filter drives a \texttt{delete}, cause unintended and silent changes to the data.

We present \textbf{MongoBench} and study four questions:
\textbf{RQ1} (capability) how accurately do LLMs predict match sets?
\textbf{RQ2} (failure structure) what makes a query hard---its semantics, or its surface complexity?
\textbf{RQ3} (source bias) when models both propose and answer items, is there an own-item advantage?
\textbf{RQ4} (mitigation) how much do rule injection, per-document decomposition, and engine delegation help, and is delegation safe?

Our contributions are:
(C1)~MongoBench, \nTotal{} items that frame consequence prediction as an exact, engine-checkable task over composed MongoDB footguns (Sec.~\ref{sec:bench});
(C2)~an evaluation of four current models including an open-weight reasoning model, with repeated runs and paired tests, plus a failure-mode analysis showing errors are better predicted by semantic pitfall type than by document count or the number of composed rules (Sec.~\ref{sec:results});
(C3)~evidence that consequence prediction is distinct from both producing a query (models mispredict 10--31\% of items they wrote) and obtaining a final answer (a model that alters query semantics while delegating) (Secs.~\ref{sec:results},~\ref{sec:mit});
(C4)~a mitigation study yielding concrete guidance for agent builders (Sec.~\ref{sec:mit}).

\section{Related Work}
\label{sec:related}

\textbf{Query understanding.}
Rahaman et al.~\cite{rahaman2025sql} probe LLM understanding of SQL via tasks such as explanation and equivalence; DBLifeBench~\cite{dblifebench-2026} spans the relational database lifecycle.
MongoBench targets the document model, whose matching rules differ sharply from SQL (implicit array traversal, \texttt{null}/missing conflation, order-sensitive embedded-document equality), and asks for an exact executable answer rather than a description.

\textbf{Query generation.}
Text-to-SQL benchmarks such as Spider~\cite{spider-2018} and BIRD~\cite{bird-2023} score generated queries by execution; SM3-Text-to-Query~\cite{sm3-2024} and Text-to-NoSQL~\cite{text2nosql-2024} extend generation to MongoDB.
A model can pass these while misjudging what its query matches---precisely the gap MongoBench measures.

\textbf{Execution reasoning.}
CRUXEval~\cite{cruxeval-2024} asks models to predict the output of short Python functions, and HumanEval~\cite{humaneval-2021} popularized execution-based checking of generated code.
MongoBench applies output prediction to a \emph{declarative} query language whose semantics are fixed by an engine rather than by a general-purpose interpreter, so there is a single authoritative answer and no hidden program state.
This difference matters for agents: an imperative program's effect is usually read off by running it, whereas a declarative filter is often reasoned about \emph{before} execution---exactly the setting where a mispredicted match set can turn into unintended data loss.
General constraint- and instruction-following benchmarks~\cite{constraintbench,cfbench,dibench} evaluate whether a model obeys stated requirements, not whether it predicts the effect of an action it emits.
To our knowledge MongoBench is the first benchmark of LLM comprehension of document-store query-matching semantics.

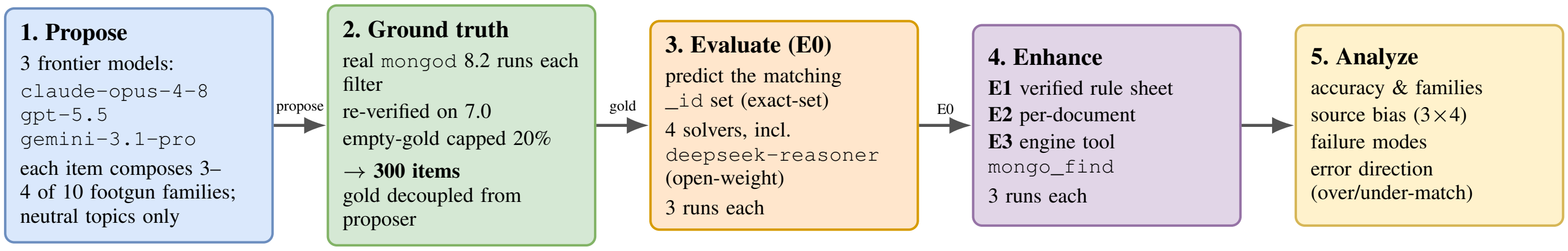
\begin{figure*}[t]
\centering
\begin{tikzpicture}[font=\scriptsize,>=Latex,
  box/.style={rectangle,rounded corners=3pt,draw,line width=.8pt,minimum height=2.35cm,text width=2.78cm,align=left,inner sep=5pt,anchor=north},
  fl/.style={->,line width=1pt,gray!65!black}]
\node[box,fill=bluF,draw=bluB] (p) {\hd{1.\ Propose}3 frontier models:\\[1pt]\mn{claude-opus-4-8}\\\mn{gpt-5.5}\\\mn{gemini-3.1-pro}\\[2pt]each item composes 3--4 of 10 footgun families; neutral topics only};
\node[box,fill=grnF,draw=grnB,right=0.62cm of p] (g) {\hd{2.\ Ground truth}real \mn{mongod} 8.2 runs each filter\\[1pt]re-verified on 7.0\\[1pt]empty-gold capped 20\%\\[3pt]\textbf{$\rightarrow$ 300 items}\\{\scriptsize gold decoupled from proposer}};
\node[box,fill=orgF,draw=orgB,right=0.62cm of g] (e) {\hd{3.\ Evaluate (E0)}predict the matching\\\mn{\_id} set (exact-set)\\[2pt]4 solvers, incl.\\\mn{deepseek-reasoner}\\{\scriptsize(open-weight)}\\[2pt]3 runs each};
\node[box,fill=purF,draw=purB,right=0.62cm of e] (m) {\hd{4.\ Enhance}\textbf{E1} verified rule sheet\\[1pt]\textbf{E2} per-document\\[1pt]\textbf{E3} engine tool \mn{mongo\_find}\\[2pt]{\scriptsize 3 runs each}};
\node[box,fill=yelF,draw=yelB,right=0.62cm of m] (a) {\hd{5.\ Analyze}accuracy \& families\\[1pt]source bias (3$\times$4)\\[1pt]failure modes\\[1pt]error direction\\{\scriptsize(over/under-match)}};
\draw[fl] (p.east)--node[above,black,font=\tiny]{propose}(p.east-|g.west);
\draw[fl] (g.east)--node[above,black,font=\tiny]{gold}(g.east-|e.west);
\draw[fl] (e.east)--node[above,black,font=\tiny]{E0}(e.east-|m.west);
\draw[fl] (m.east)--(m.east-|a.west);
\end{tikzpicture}
\caption{The MongoBench pipeline. Three vendors' frontier models propose items that compose three or four documented MongoDB matching pitfalls; a real \mongod server computes ground truth, decoupled from the proposer; four models (including one open-weight) are evaluated; and three interventions are measured.}
\label{fig:pipeline}
\end{figure*}

\section{The MongoBench Benchmark}
\label{sec:bench}

Fig.~\ref{fig:pipeline} gives an overview.
\textbf{Task and metrics.}
Each item is a collection $D$ (JSON documents with integer \texttt{\_id}s) and a \find filter $Q$; the model returns the \texttt{\_id} set that $Q$ matches.
The primary metric is \emph{exact-set accuracy}: correct only if the predicted set equals the gold set---a preview that misses one document is wrong.
We also report per-document accuracy and Jaccard similarity to separate matching judgment from set construction.

\textbf{Footgun families.}
Items draw on ten documented behaviors, each a rule where MongoDB diverges from a na\"ive reading:
(1)~\emph{array-implicit matching}---a scalar condition \texttt{\{f:v\}} also matches a document whose \texttt{f} is an array containing \texttt{v};
(2)~\emph{implicit-AND vs.\ \texttt{\$elemMatch}}---two conditions on one array field may be satisfied by two \emph{different} elements unless \texttt{\$elemMatch} forces a single element;
(3)~\emph{\texttt{null}/missing}---\texttt{\{f:null\}} matches documents where \texttt{f} is \texttt{null}, missing, or an array containing \texttt{null};
(4)~\emph{negation over arrays}---\texttt{\$ne}/\texttt{\$nin} exclude a document if \emph{any} array element matches;
(5)~\emph{type bracketing}---comparisons never cross BSON types, so \texttt{\{f:\{\$gt:5\}\}} never matches a string;
(6)~\emph{comparison vs.\ \texttt{null}}---\texttt{\$gt}/\texttt{\$lt} against \texttt{null} match nothing, while \texttt{\$gte}/\texttt{\$lte} match \texttt{null} and missing;
(7)~\emph{embedded-document equality}---a whole-document value requires the exact field set in the exact order;
(8)~\emph{dot-notation into arrays}---\texttt{a.b} traverses an array of sub-documents, and \texttt{a.0} indexes positionally;
(9)~\emph{nested arrays}---implicit matching descends only one array level; and
(10)~\emph{\texttt{\$size}/\texttt{\$all}}---exact length and containment, including \texttt{\$all} with \texttt{\$elemMatch}.
Fig.~\ref{fig:ex} shows two in isolation.

\textbf{Composition.}
Every item composes \emph{three or four} of these families at once: a numbered specification names only the neutral topics involved (e.g.\ ``negation when the field is an array''), and the proposer must design an item whose correct answer depends on all of them.
Proposers never see rules, answers, or other items.
Collections hold \meanDocs{} documents on average.
We exclude single-rule items to target composition, which is the open question.

\begin{figure}[t]
\small
\fbox{\parbox{0.93\columnwidth}{
\textbf{(a) range vs.\ \texttt{null}.}\\
$D$: \texttt{\{\_id:1,f:5\}, \{\_id:2,f:null\}, \{\_id:3\}}\\
$Q$: \texttt{\{f:\{\$gt:null\}\}}\quad Gold: $\emptyset$\\
A model reading \texttt{\$gt:null} as ``$f$ is not null'' predicts $\{1\}$.

\medskip
\textbf{(b) embedded-document order.}\\
$D$: \texttt{\{\_id:1,a:\{b:1,c:2\}\}, \{\_id:2,a:\{c:2,b:1\}\},}\\
\phantom{$D$: }\texttt{\{\_id:3,a:\{b:1,c:2,d:0\}\}}\\
$Q$: \texttt{\{a:\{b:1,c:2\}\}}\quad Gold: $\{1\}$\\
Whole-document equality is field-order and field-set sensitive, so 2 and 3 do not match; treating it as a subset or order-insensitive predicts $\{1,2,3\}$.
}}
\caption{Two footguns (verified on \mongod). Both read naturally but match counter-intuitively; MongoBench items compose several such rules.}
\label{fig:ex}
\end{figure}

\textbf{Proposers and symmetry.}
The \nTotal{} items are proposed in equal shares (\nProposer{} each) by \texttt{opus-4-8}, \texttt{gpt-5.5}, and \texttt{gemini-3.1-pro}---the same three models evaluated---so item source is an explicit factor and any own-item advantage is measurable (RQ3).

\textbf{Ground truth and validity.}
Gold match sets are computed by executing each filter on a real \mongod server (v8.2.9)~\cite{mongodb-manual}, independent of the proposer, following execution-based evaluation practice~\cite{humaneval-2021,bird-2023}.
All gold sets were recomputed on \mongod 7.0.34 and agree exactly, ruling out version-specific quirks.
We use \mongod directly rather than an in-process emulator.
Items are kept only on model-independent validity criteria (parses as strict JSON, executes, in-range document count, not a duplicate).
Because a composed filter containing, e.g., a \texttt{\$gt:null} conjunct trivially matches nothing, empty gold sets are capped at 20\% per proposer; the resulting always-empty baseline scores \emptyBase\%.

\section{Experimental Setup}
\label{sec:setup}

\textbf{Models.}
We evaluate frontier models from three vendors---\texttt{claude-opus-4-8}~\cite{anthropic-claude}, \texttt{gpt-5.5}~\cite{openai-gpt}, and \texttt{gemini-3.1-pro-preview}~\cite{gemini-team} (hereafter \texttt{gemini-3.1-pro})---and one open-weight reasoning model accessed through DeepSeek's \texttt{deepseek-reasoner} API endpoint~\cite{deepseek-r1} (we name it by that API identifier). All were accessed via vendor APIs in late 2026.
All three run at \emph{medium} reasoning effort; \texttt{deepseek-reasoner} reasons by default (its effort is not configurable), which we note when comparing.
We report model versions and settings rather than equalize them.

\textbf{Protocol.}
The baseline prompt (E0) shows documents and filter as JSON and asks for a final line \texttt{ANSWER: [ids]}; no model produced an unparseable answer in any run.
The main experiment is repeated three times per model; we report mean\,$\pm$\,std, and for paired model comparisons take each item's majority outcome and apply McNemar's exact test~\cite{mcnemar1947} with Holm correction~\cite{holm1979}.
Enhancement conditions (Sec.~\ref{sec:mit}) are each run three times.

\section{Results}
\label{sec:results}

\begin{table}[t]
\centering
\caption{Baseline accuracy (E0, 3-run mean\,$\pm$\,std, $n{=}\nTotal$). Accuracy is the 3-run mean; per-document accuracy and Jaccard are from one run. Tables abbreviate \texttt{deepseek-reasoner} as \texttt{deepseek}.}
\label{tab:main}
\small
\input{gen3/tab_main.tex}
\end{table}

\textbf{RQ1: capability.}
Table~\ref{tab:main} reports baseline accuracy.
\texttt{gpt-5.5} leads at \accGpt\%, \texttt{opus-4-8} reaches \accOpus\%, and \texttt{gemini-3.1-pro} and \texttt{deepseek-reasoner} trail at \accGem\% and \accDs\%---a \gapTop-point spread among current models on a task with a \emptyBase\% trivial baseline.
Reasoning budgets are not matched across vendors (\texttt{deepseek-reasoner}'s effort is not configurable), so this spread reflects capability \emph{and} effort together, not capability alone (Sec.~\ref{sec:threats}); the per-item and safety findings below do not depend on the ranking.
Run-to-run std is below one point.
Per-document accuracy is far higher than exact-set accuracy for every model (e.g.\ \texttt{gemini-3.1-pro} \perdocGem\% per-doc vs.\ \exactGemRun\% exact-set), so a weak model typically misjudges only one or two documents per item---in production that single misjudged document is the silent error.
All pairwise accuracy differences are significant (McNemar's exact test, Holm-corrected $p<0.05$) except \texttt{deepseek-reasoner} vs.\ \texttt{gemini-3.1-pro} ($p{=}\dsGemP$).
Only \nAllWrong{} items are missed by all four models, so these are model-dependent blind spots rather than universally impossible items.
Composition depth barely matters: items joining four families are no harder than three for any model (\texttt{gemini-3.1-pro} scores \byThreeGem\% and \byFourGem\%), consistent with the failure analysis below: difficulty comes from the \emph{kind} of rule, not the number stacked.

\begin{figure}[t]
\centering
\includegraphics[width=\columnwidth]{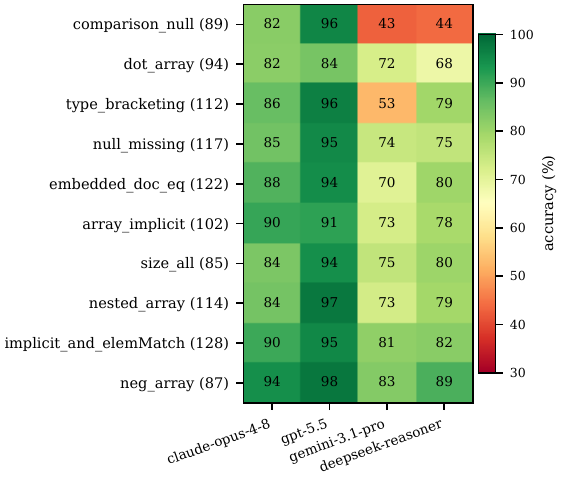}
\caption{Accuracy (\%) by footgun family (majority of 3 runs); an item counts toward every family it composes. Families sorted by mean accuracy.}
\label{fig:family}
\end{figure}

\textbf{RQ2: what makes a query hard.}
Fig.~\ref{fig:family} breaks accuracy down by family.
Comparison against \texttt{null} is hardest---\texttt{gemini-3.1-pro} \compNullGem\% and \texttt{deepseek-reasoner} \compNullDs\% vs.\ \texttt{gpt-5.5} \compNullGpt\%---followed by type bracketing: the rules that contradict the ``SQL-like'' intuition that a comparison with \texttt{null} means ``is not null'' and that cross-type values are ordered.
To ask what drives difficulty, we label each item with structural features and relate them to the number of models that fail it.
The strongest signal is an empty gold set (\failEmptyPres\% of such items are missed by at least one model vs.\ \failEmptyAbs\%, Fisher \failEmptyP); being a property of the answer, we treat it as descriptive and exclude it (and result-set size) from the predictor below. Among pre-execution features, a range comparison (\failRangePres\% vs.\ \failRangeAbs\%) and a \texttt{null} literal are most associated with failure.
The surface measures we tested are not: accuracy does not fall as the collection grows (Pearson \corrNdocs; Table~\ref{tab:ndocs} shows no monotonic trend over the 5--12 documents per item) and composing four families is no harder than three (standardized coefficient \coefKfour).
A logistic model over the pre-execution features alone predicts failure at \logregHeldout\% held-out accuracy (leave-one-out cross-validation; \logregPreexec\% in-sample) against a \logregBase\% base rate; Table~\ref{tab:fail} lists its strongest coefficients.
Within the features we measured, difficulty tracks \emph{which} semantic pitfall a query exercises more than its document count or the number of composed rules---evidence of genuine misunderstanding, though these remain associations (items compose several families) rather than isolated causes.
(Queries with \texttt{\$ne}/\texttt{\$nin} are, interestingly, \emph{easier}: their larger, more intuitive result sets are harder to get wrong.)

\begin{table}[t]
\centering
\caption{Strongest predictors that an item is failed by at least one model (standardized logistic-regression coefficients; $+$ = harder).}
\label{tab:fail}
\small
\input{gen3/tab_fail.tex}
\end{table}

\textbf{Error direction.}
When a model is wrong, does it over- or under-predict the match set?
Table~\ref{tab:errdir} classifies each error as an over-match (predicted a strict superset of gold), under-match (strict subset), or incomparable set.
Errors lean toward over-matching overall (\overPooled\% vs.\ \underPooled\%), but the direction is \emph{model-specific}: \texttt{gpt-5.5} and \texttt{gemini-3.1-pro} over-match on $\sim$\overGpt\% of their errors, whereas \texttt{opus-4-8} under-matches on \underOpus\%.
This matters for safety: an over-matching preview overstates a \texttt{delete}'s blast radius (false alarm), while an under-matching one understates it---silently leaving matched documents unaccounted for.
Because no model errs consistently in the ``safe'' direction, an agent cannot assume its mispredictions are conservative; this is a further argument for delegating to the engine (Sec.~\ref{sec:mit}).

\begin{table}[t]
\centering
\caption{Direction of errors (E0, among wrong items). Over-match = prediction is a strict superset of gold; under-match = strict subset.}
\label{tab:errdir}
\small
\setlength{\tabcolsep}{5pt}
\input{gen3/tab_errdir.tex}
\end{table}

\begin{table}[t]
\centering
\caption{Accuracy (\%) by item source (rows) and solver (columns), majority of 3 runs. Underlined: a model answering its own items. Only these three models propose items.}
\label{tab:source}
\scriptsize
\setlength{\tabcolsep}{3pt}
\input{gen3/tab_source.tex}
\end{table}

\textbf{RQ3: proposing an item is not solving it.}
Table~\ref{tab:source} crosses item source with solver.
Each proposer fails on items it wrote itself: \texttt{opus-4-8}, \texttt{gpt-5.5}, and \texttt{gemini-3.1-pro} get \ownWrongOpus\%, \ownWrongGpt\%, and \ownWrongGem\% of their own items wrong, so generating a query that exercises a footgun does not imply predicting its result---producing a query and predicting its consequence are distinct abilities.
On every source, the best solver is a model \emph{other} than the item's author (\texttt{gpt-5.5} on \texttt{opus-4-8}'s and \texttt{gemini-3.1-pro}'s items, \texttt{opus-4-8} on \texttt{gpt-5.5}'s), though the sources differ in difficulty (see Sec.~\ref{sec:threats}).

\section{Mitigation and the Delegation Gap}
\label{sec:mit}

\begin{table}[t]
\centering
\caption{Enhancement accuracy (\%), 3-run mean. E1: prepend a verified 10-rule sheet. E2: force per-document judgment. E3: let the model run the filter via a \texttt{mongo\_find} tool.}
\label{tab:enh}
\small
\resizebox{\columnwidth}{!}{\input{gen3/tab_enh.tex}}
\end{table}

We compare three interventions against E0 (Table~\ref{tab:enh}), each run three times.

\begin{figure}[t]
\centering
\includegraphics[width=\columnwidth]{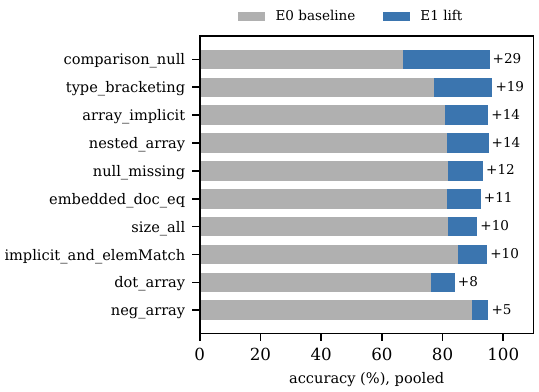}
\caption{Per-family accuracy before (E0) and after (E1) rule injection, pooled over models. Large lift is consistent with a knowledge gap the rule closes; small lift, with difficulty applying a rule the model has now been given.}
\label{fig:lift}
\end{figure}

\textbf{Rule injection (E1)} prepends a one-line rule per family (Fig.~\ref{fig:rules}), each statement verified against \mongod.
It helps unevenly: \texttt{gemini-3.1-pro} rises to \eoneGem\% and \texttt{deepseek-reasoner} to \eoneDs\%, while \texttt{gpt-5.5} is unchanged (\eoneGpt\% vs.\ \ezeroGpt\%; the small drop is not significant, paired $p\!\approx\!0.24$).
Per family (Fig.~\ref{fig:lift}), the lift is largest where models lacked the rule---comparison vs.\ \texttt{null} (\liftCompNull), type bracketing---and smaller for dot-notation into arrays (\liftDotArray{} points).
This pattern is consistent with a \emph{knowledge gap} in the first case and difficulty \emph{applying} a known rule in the second.
Rule injection thus appears to patch missing knowledge more than faulty application, and does not help every model.

\begin{figure}[t]
\scriptsize
\fbox{\parbox{0.93\columnwidth}{\ttfamily
Array-implicit: \{f:v\} matches f==v OR an array elem ==v\\
Implicit-AND: \{f:\{\$gt:a,\$lt:b\}\} diff elems ok; \$elemMatch=one\\
null/missing: \{f:null\} matches null, missing, or [..null..]\\
Compare vs null: \$gt/\$lt:null match NOTHING;\\
\phantom{xx}\$gte/\$lte:null match null/missing\\
Type bracket: comparisons stay within one BSON type\\
\$ne/\$nin: exclude doc if ANY array elem matches\\
Embedded-doc eq: requires exact field set AND order\\
Dot notation: a.b into array elems; a.0 positional;\\
\phantom{xx}does not reach arrays nested in arrays\\
Nested arrays: implicit match descends one level only\\
\$size/\$all: exact length; \$all=contains all (impl.)
}}
\caption{The E1 rule sheet (one line per family, abbreviated; each rule was verified against \mongod before use, full text released). This paragraph is prepended to the E0 prompt.}
\label{fig:rules}
\end{figure}

\textbf{Decomposition (E2)}, judging each document before assembling the set, has a small, inconsistent effect (it raises \texttt{opus-4-8} by about four points, to \etwoOpus\%, but leaves the other three within two) and is not a reliable substitute.

\textbf{Engine delegation (E3)} gives the model a tool that runs a filter on the collection and returns the matching \texttt{\_id}s (up to ten calls).
\texttt{opus-4-8} and \texttt{gpt-5.5} reach 100\% and \texttt{deepseek-reasoner} \dsEthreeVal\%.
But delegation is not automatically safe.
We canonicalize each executed filter against the intended one, treating order-irrelevant rewrites (reordered top-level conditions or operators) as preserving and order-relevant ones (an embedded document's field order, or a changed key) as altering; this check is conservative, as it may flag rewrites that are in fact equivalent.
Across all four models and all three runs, only \texttt{gemini-3.1-pro} produced items on which it called the tool yet no executed filter passed this check---\gemAlteredPctMean\% of items on average (\gemAlteredPctLo--\gemAlteredPctHi\% across runs); the other three passed on every item.
On about \gemAlteredWrongMean{} of those items per run (range \gemAlteredWrongLo--\gemAlteredWrongHi{}), \texttt{gemini-3.1-pro}'s final answer was also wrong. Fig.~\ref{fig:deleg} illustrates a concrete alteration that changes the matching result: asked to run \texttt{\{...,\ "v.0":\{\$exists:true\}\}}, \texttt{gemini-3.1-pro} quoted the positional-index key as the literal string \texttt{"'v.0'"}, so the executed filter tested a field literally named \texttt{v.0} instead of the first array element; eight tool calls later it still returned the empty set instead of the correct \texttt{\{1,2\}}.
The engine answered faithfully---but to a different query than intended.
Adversaries can exploit this gap to make agents issue backend queries that slip past access controls~\cite{fragfuse}; here it opens \emph{benignly}, with no adversary.

\begin{figure}[t]
\scriptsize
\fbox{\parbox{0.93\columnwidth}{\ttfamily
intended:\ \{a:\{b:1,c:2\}, v:\{\$gt:3,\$lt:7\}, v.0:\{\$exists:true\}\}\\
executed: \{a:\{b:1,c:2\}, v:\{\$gt:3,\$lt:7\}, 'v.0':\{\$exists:true\}\}\\
gold \{1,2\}\ \ returned \{\}\ \ (8 tool calls, never corrected)
}}
\caption{Engine delegation can still fail: \texttt{gemini-3.1-pro} alters the query's meaning (quoting a positional-index key) before executing it, so a faithful engine returns the wrong set.}
\label{fig:deleg}
\end{figure}

\textbf{Guidance for agent builders.}
(1)~For previewing or validating a query, execute the \emph{original} filter against the engine and return its result verbatim---do not let the model reconcile the engine's answer with its own prediction.
(2)~If in-context reasoning is unavoidable, include a verified rule sheet for the declarative footguns, but expect residual errors on execution-style rules such as dot-notation.
(3)~Do not rely on per-document decomposition as a safeguard.

\begin{table}[t]
\centering
\caption{Accuracy (\%) by collection size (majority of 3 runs). Accuracy shows no monotonic decline as the collection grows.}
\label{tab:ndocs}
\small
\setlength{\tabcolsep}{4pt}
\resizebox{\columnwidth}{!}{\input{gen3/tab_ndocs.tex}}
\end{table}

\section{Threats to Validity}
\label{sec:threats}

\textbf{Model coverage and budget.} We evaluate four models at their default or medium reasoning settings; reasoning budgets are not matched across vendors, and \texttt{deepseek-reasoner}'s effort is not configurable, so cross-model gaps conflate capability and effort.
\textbf{LLM-proposed items and source difficulty.} Items are written by the evaluated frontier models, with no human-written control; gold never depends on a proposer, and we report accuracy stratified by source (RQ3). The three sources differ in difficulty (empty-gold items number 3, 20, and 19 across proposers), so own-item effects cannot be fully separated from source difficulty and we do not claim a difficulty-matched null; a matched, human-controlled source\,$\times$\,solver analysis is future work.
\textbf{Rule sheet as upper bound.} The E1 rules name the same ten footgun families the items are built from, so E1 measures accuracy when the relevant rule is supplied verbatim---an upper bound on in-context help, not what an agent would have unprompted.
\textbf{Saturation.} Only four items are missed by all four models, so the hardest stratum is near the ceiling of what these models cannot do; a harder item pool may be needed to separate future models.
\textbf{Scope.} We cover \find filters, not projections, updates, or aggregation pipelines; extending the consequence-prediction framing to those is future work.

\section{Conclusion}
\label{sec:conclusion}

Evaluating an agent that acts on a stateful system should not stop at ``can it produce an action'' or ``can it get an answer''; it must also ask whether the agent knows what its action will do, and whether it preserves that action's meaning at the point of execution.
MongoBench measures the first for MongoDB \find filters and surfaces a concrete instance of the second.
All four current models misjudge composed matching footguns---failures tracking semantic pitfall type more than document count or the number of composed rules---rule sheets help unevenly, and even engine delegation can be undone by a model that silently rewrites the query.
The same question applies to any setting where what a model emits is itself an action whose effect a deterministic system can adjudicate.
A code edit has a test suite; an access-control rule has a policy evaluator; a configuration change has the system it configures.
In each case an agent can generate the artifact and can run the checker, yet, as here, neither establishes that it predicted what the artifact would do, nor that the artifact handed to the checker still meant what the agent intended.
Where such a ground-truth system is cheap to call, the engineering answer is to call it and use its verdict verbatim; the open evaluation question is how well agents know, in advance, what their own actions will do.
We release MongoBench---the items, the \mongod oracle, and the evaluation code with all prompts at \repoURL---as a template for building such consequence-prediction benchmarks in other executable domains.

\section*{Acknowledgment}
As detailed in Secs.~\ref{sec:bench} and~\ref{sec:results}, large language models are both the subjects of this evaluation and the proposers of the composed items, whose answers were computed by an independent \mongod oracle and checked for source bias. AI tools were also used to assist with implementation and writing. All methodological choices, analyses, and conclusions are the authors'.

\bibliographystyle{IEEEtran}
\bibliography{references}

\end{document}

%% file: gen3/numbers.tex
\newcommand{\nTotal}{300}
\newcommand{\emptyBase}{14.0}
\newcommand{\meanDocs}{7.9}
\newcommand{\nProposer}{100}
\newcommand{\accOpus}{85.1}

\newcommand{\accGpt}{92.1}

\newcommand{\accGem}{70.1}

\newcommand{\accDs}{72.3}

\newcommand{\gapTop}{22}

\newcommand{\ownWrongOpus}{11}

\newcommand{\ownWrongGpt}{10}

\newcommand{\ownWrongGem}{31}
\newcommand{\ezeroGpt}{92.1}
\newcommand{\eoneGpt}{91.2}

\newcommand{\etwoOpus}{89.3}

\newcommand{\eoneDs}{93.3}

\newcommand{\eoneGem}{95.3}

\newcommand{\liftCompNull}{+29}
\newcommand{\liftDotArray}{+8}
\newcommand{\gemAlteredPctMean}{28}
\newcommand{\gemAlteredPctLo}{27}
\newcommand{\gemAlteredPctHi}{30}

\newcommand{\gemAlteredWrongMean}{22}
\newcommand{\gemAlteredWrongLo}{16}
\newcommand{\gemAlteredWrongHi}{29}

\newcommand{\dsEthreeVal}{99.7}
\newcommand{\byThreeGem}{70}
\newcommand{\byFourGem}{70}
\newcommand{\perdocGem}{92.2}
\newcommand{\exactGemRun}{70.7}
\newcommand{\dsGemP}{0.06}

\newcommand{\compNullGem}{43}
\newcommand{\compNullDs}{44}
\newcommand{\compNullGpt}{96}
\newcommand{\nAllWrong}{4}
\newcommand{\failEmptyPres}{93}
\newcommand{\failEmptyAbs}{37}
\newcommand{\failEmptyP}{$3.0{\times}10^{-12}$}
\newcommand{\failRangePres}{50}
\newcommand{\failRangeAbs}{19}

\newcommand{\corrNdocs}{-0.03}

\newcommand{\logregBase}{55}
\newcommand{\logregPreexec}{67}
\newcommand{\logregHeldout}{62}

\newcommand{\coefKfour}{-0.12}
\newcommand{\overPooled}{63}
\newcommand{\underPooled}{33}
\newcommand{\overGpt}{82}

\newcommand{\underOpus}{63}

%% file: gen3/tab_main.tex
\begin{tabular}{lcc}
\toprule
Model & Accuracy (\%) & per-doc / Jaccard \\
\midrule
\texttt{gpt-5.5} & \textbf{92.1$\pm$0.5} & 98.7 / 96.4 \\
\texttt{opus-4-8} & 85.1$\pm$0.7 & 96.8 / 92.8 \\
\texttt{deepseek} & 72.3$\pm$0.9 & 93.3 / 82.5 \\
\texttt{gemini-3.1-pro} & 70.1$\pm$0.7 & 92.2 / 80.4 \\
\midrule
Always-empty & 14.0 & -- \\
\bottomrule
\end{tabular}

%% file: gen3/tab_fail.tex
\begin{tabular}{lcc}
\toprule
Query feature & std.\ coef. & effect \\
\midrule
range comparison (\texttt{\$gt}/\texttt{\$lt}) & $+0.73$ & harder \\
negation (\texttt{\$ne}/\texttt{\$nin}) & $-0.53$ & easier \\
\texttt{\$elemMatch} & $-0.48$ & easier \\
\texttt{null} literal & $+0.46$ & harder \\
dot-notation path & $+0.46$ & harder \\
\texttt{\$exists} & $-0.18$ & easier \\
\bottomrule
\end{tabular}

%% file: gen3/tab_errdir.tex
\begin{tabular}{lccc}
\toprule
Model & over-match & under-match & incomp. \\
\midrule
\texttt{opus-4-8} & 33\% & 63\% & 5\% \\
\texttt{gpt-5.5} & 82\% & 14\% & 5\% \\
\texttt{gemini-3.1-pro} & 81\% & 17\% & 2\% \\
\texttt{deepseek} & 55\% & 39\% & 6\% \\
\bottomrule
\end{tabular}

%% file: gen3/tab_source.tex
\begin{tabular}{lcccc}
\toprule
Proposer ($n{=}\nProposer$) & \texttt{opus-4-8} & \texttt{gpt-5.5} & \texttt{gemini-3.1-pro} & \texttt{deepseek} \\
\midrule
\texttt{opus-4-8} & \underline{89} & 96 & 72 & 83 \\
\texttt{gpt-5.5} & 91 & \underline{90} & 69 & 66 \\
\texttt{gemini-3.1-pro} & 78 & 96 & \underline{69} & 78 \\
\bottomrule
\end{tabular}

%% file: gen3/tab_enh.tex
\begin{tabular}{lcccc}
\toprule
Model & E0 & E1 rules & E2 per-doc & E3 engine \\
\midrule
\texttt{gpt-5.5} & 92.1 & 91.2 & 92.0 & \textbf{100.0} \\
\texttt{opus-4-8} & 85.1 & 93.2 & 89.3 & \textbf{100.0} \\
\texttt{deepseek} & 72.3 & 93.3 & 74.1 & \textbf{99.7} \\
\texttt{gemini-3.1-pro} & 70.1 & \textbf{95.3} & 72.2 & 92.1 \\
\bottomrule
\end{tabular}

%% file: gen3/tab_ndocs.tex
\begin{tabular}{lcccc}
\toprule
\#docs ($n$) & \texttt{opus-4-8} & \texttt{gpt-5.5} & \texttt{gemini-3.1-pro} & \texttt{deepseek} \\
\midrule
5 (34) & 74 & 91 & 74 & 76 \\
6--7 (75) & 81 & 96 & 69 & 80 \\
8 (85) & 91 & 92 & 73 & 71 \\
9--12 (106) & 90 & 95 & 67 & 76 \\
\bottomrule
\end{tabular}